\documentclass[a4paper,twoside]{article}
\newcommand{\affil}[1]{$^{\rm #1}$}
\usepackage[authoryear]{natbib}
\bibpunct{(}{)}{;}{a}{}{,}
\usepackage{graphicx}
\date{} 
\title{\large\bf\flushleft Feature Detection in Radio Astronomy using the Circle Hough Transform}
\author{\parbox{\textwidth}{\flushleft
\vspace{-0.5cm}
{\it Christopher Hollitt\affil{A,C} and Melanie Johnston-Hollitt\affil{B}}\\
\vspace{0.4cm}
{\small \affil{A}\,School of Engineering and Computer Science\\
    Victoria University of Wellington\\
    PO Box 600, Wellington 6140}\\
{\small \affil{B}\,School of Chemical and Physical Sciences\\
    Victoria University of Wellington\\
    PO Box 600, Wellington 6140}\\
{\small \affil{C}\,Email: chollitt@ieee.org}}}
\begin{document}
\twocolumn[
\begin{changemargin}{.8cm}{.5cm}
\begin{minipage}{.9\textwidth}
\vspace{-1cm}
\maketitle
%
%
\small{\bf Abstract:}
While automatic detection of point sources in astronomical images has
experienced a great degree of success, less effort has been directed towards
the detection of extended and low-surface brightness features. At present,
existing telescopes still rely on human expertise to reduce the raw data to
usable images and then to analyse the images for non-pointlike objects.
However, the next generation of radio telescopes will generate unprecedented
volumes of data making manual data reduction and object extraction infeasible.
Without developing new methods of automatic detection for extended and diffuse
objects such as supernova remnants, bent-tailed galaxies, radio relics and
halos, a wealth of scientifically important results will not be uncovered. In
this paper we explore the response of the Circle Hough Transform to a
representative sample of different extended circular or arc-like astronomical
objects. We also examine the response of the Circle Hough Transform to input
images containing noise alone and inputs including point sources.


\medskip
\medskip
\end{minipage}
\end{changemargin}
]
\small

\section{Introduction}
\label{sec:Introduction}

The radio sky is rich in scientifically important objects containing circular
or arc-like structures. These objects range in physical size from the shells of
dying stars to shocks created in merging clusters of galaxies. On the small
scale, objects such as supernova remnants stereotypically contain complete or
partial circular shells, or in cases when their symmetry is distorted, both
\citep{mineshige90}. Arc-like features are also found in Active Galactic Nuclei
(AGN), either in the overall source shape as in the case of bent-tail radio
galaxies \citep{mjh04,mao2009}, or as features observed within the lobes of the
AGN themselves \citep{feain11}. Finally, giant Mpc-scaled radio relics, which
occur on the edges of merging clusters, are found as single or double arc-like
regions \citep{mjh03, Bonafede09}. In addition to their arc-like morphology,
these objects are often low in surface brightness and exhibit breaks or gaps
which make them hard to detect in an automated fashion. This is particularly
true of techniques that rely on either high signal-to-noise or continuous
edges in order to define the full extent of the source.

In order to reliably detect broken, low-surface brightness sources we require
algorithms which do not rely solely on the behaviour of individual pixels, but
exploit additional non-local information. The advantages of such algorithms are
that they are robust in the presence of non-connectedness and noise. Examples of
such techniques include template matching, wavelet and related integral
transforms and Hough transforms. As the Hough transform is known to be robust to
the presence of partial or occluded target objects and to noise
\citep{Davies_1988}, it is a natural choice to explore for the automatic
detection of circular features in astronomical data.

There has been some previous use of the Hough transform in astronomy. It has
been used to remove contamination from optical surveys \citep{Storkey_2004}, to
characterise echelle orders \citep{Ballester94, Ballester_1996}, and to
characterise craters on the moon \citep{Jahn_1994}. However, it has not been
previously applied for either source detection or characterisation in radio
astronomy. The increased resolution and field of view that will become routine
when using instruments like the Square Kilometre Array (SKA)
\citep{Schilizzi08} and its precursors such as the Australian SKA Pathfinder
(ASKAP) and the Murchison Widefield Array (MWA), force a paradigm shift towards
greater reliance on automated data extraction techniques such as that described
here.
 
In this work we concentrate on the detection problem; that of locating circular
regions of radio emission that are unlikely to have arisen from statistical
fluctuations in image noise. We also believe that the techniques described in
this paper could contribute to the related characterization problem, where
morphological features of the uncovered emission are accurately determined. As
a result in this paper we have concentrated on detecting circular structures,
rather than the accuracy or precision with which they are described.

In this paper we present a description of the circle Hough transform (CHT) in
Section \ref{sec:Method}, followed by a detailed discussion of the applications
of the CHT as a detection algorithm for use in Radio Astronomy in Section
\ref{sec:Results}. Section \ref{sec:Discussion} presents a discussion of
computational issues and response to image artifacts and presents a suggested
framework for the practical implementation of the CHT for next generation radio
telescopes. We present a summary of the work and final conclusions in Section
\ref{sec:Conclusion}. We have adopted a standard set of cosmological parameters
throughout with $H_0 = 73 \mathrm{~km\,s}^{-1}\mathrm{Mpc}^{-1}$, $\Omega_m =
0.27$ and ${\Omega}_{\Lambda}=0.73$.

\section{Circle Hough Transforms}
\label{sec:Method}
 
The family of algorithms known collectively as Hough transforms are capable of
finding and characterising a variety of geometrical objects in image or image
like-data \citep{Duda_1972,Kerbyson_1995}. The application of the Hough
transform for circle detection was a very early development in the field, and
has since found wide application \citep{Kimme_1975,Illingworth_1988,Rad_2003}.

In its simplest form the circle Hough transform is computationally
challenging, with both computational effort and memory consumption scaling as
$O(n^3)$ when transforming an $n \times n$ pixel image. Many variants of the CHT
algorithms have consequently been developed to improve the computational
performance of the algorithm
\citep{Davies_1988,Kerbyson_1995,Ioannou_1999,Chiu_2010}, but current
implementations of the CHT remain confined mostly to offline processing.

As is the case for all Hough transforms, the CHT seeks to map an image into a
quantised parameter space that describes the target objects. An arbitrary circle
in two dimensions can be described by
\begin{equation}
  (x-x_0)^2+(y-y_0)^2-r^2=0 \, ,
  \label{eqn:circle_equation} 
\end{equation}
where $x_0$ and $y_0$ are the x and y coordinates of the centre of the circle
and $r$ is the circle's radius. Each different three-tuple
$\mathbf{x}:=(x_0,y_0,r)$ uniquely parametrises different circles that may be
present in an image. Each point in the Hough space therefore corresponds
to a complete circle with a particular location and size. The operation of the
CHT is to map the input image into the parameter space such that the transform
has a high value for parameter tuples consistent with objects that are present
in the image.

In its common form, the Hough transform operates on each image pixel
independently. For each pixel the algorithm finds the set of parameter
tuples that would result in an object passing through the considered
pixel. That is, consideration of an image pixel $f(x_i,y_i)$ finds the
set $\mathcal{P}$ of all $(x_0,y_0,r)$ tuples that satisfy equation
\ref{eqn:circle_equation};
\begin{equation}
  \mathcal{P}=\left\{(x_0,y_0,r):(x_i-x_0)^2+(y_i-y_0)^2-r^2=0\right\}
  \label{eqn:circle_equation_point} 
\end{equation}

Each location in Hough space that corresponds to a member of $\mathcal{P}$ is
then incremented. Thus each image pixel can be regarded as casting a set of
``votes'', for those parameter tuples that are consistent with itself.
Examination of equation \ref{eqn:circle_equation_point} reveals that the set of
votes arising from a single pixel forms a right cone in Hough space
\citep{Duda_1972, Yuen_1990}.

\begin{figure} 
  \centering 
  \includegraphics[width=\columnwidth]{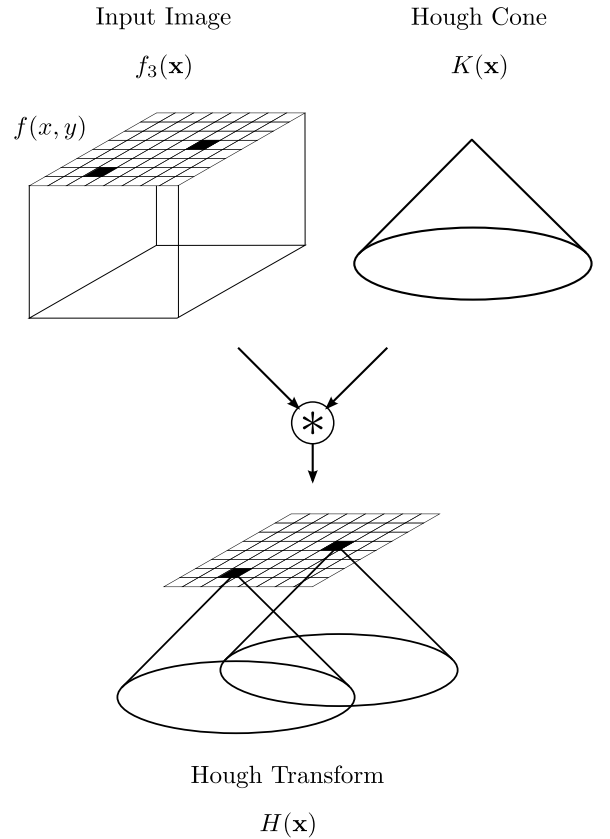}
  \caption{Formation of the Hough transform via convolution.
  The input image $f(x,y)$ is extended to three dimensions and then convolved
  with the vote distribution $K(\mathbf{x})$ to generate the Hough transform
  $H(\mathbf{x})$. For clarity we have represented the Hough cone as smooth,
  though it is discretized on the same scale as the input image.}
  \label{fig:Hough_cartoon}
\end{figure}

The vote distributions from multiple image pixels are added and produces a high
number of votes for parameter combinations that describe geometric figures that
are actually present in the image. The operation of the CHT can be conveniently
regarded as the three-dimensional convolution of the input image with the
conical vote distribution \citep{Hollitt_2009}, as illustrated in Figure
\ref{fig:Hough_cartoon}.

For our application, if we calculate the CHT of a radio image and then find the
peak in the Hough transform we will recover the radius and center of any
circular astronomical features in the input radio map.

\subsection{Detection Procedure}
\label{sec:Detection}

Peaks in the Hough transform of an arbitrary input image can arise even in the
absence of a circular region of diffuse emission. In this section we briefly
discuss the production of peaks due to noise in the input image and note
that peaks are also expected because of the presence of point sources.

Let us first consider the expected Hough peak size that would arise in Hough
space as a result of input image amplitude noise, which we expect to be normally
distributed with mean $\hat{a}$ and standard deviation $s$ (that is the input
image is $\sim\mathcal{N}\left( \hat{a}, s^2 \right)$). When searching for a
circle having radius $r$ the Hough transform adds the magnitudes of
approximately $2\pi r$ pixels. For a random input image we would therefore
expect that the distribution of magnitudes for the Hough transform would be
$\sim\mathcal{N}\left( 2 \pi r \hat{a},  2 \pi r s^2 \right)$.

One can therefore select an appropriate threshold for the source detector as $2
\pi r \hat{a} + \zeta s \sqrt{2 \pi r}$, for $\zeta$ the number of standard
deviations above the expected mean that one wishes to use. For an image a few
hundred pixels in extent, $\zeta=5$ should ensure that Hough peaks arising
solely from noise are rarely encountered. A considerably more aggressive
threshold could be used if a higher false-positive detection rate were
acceptable for a particular application.

The noise in radio images arising from surveys is unlikely to be stationary, so
the appropriate values of $\hat{a}$ and $s$ will need to be calculated locally
\citep{Huynh2011}. Note also that the analysis above would need to change
slightly if the Hough transform were used to search for structures in
polarimetric surveys, as the noise in polarisation images is Rician rather than
Gaussian.

We would also expect that the Hough transform of an arbitrary radio image will
contain a number of putative circles due to the presence of point sources.
In a noiseless image the Hough transform would find circles centered half
way between each pair of point sources. Such circles will have a diameter
approximately equal to the separation of the points sources. However in a noisy
image the circle is typically slightly displaced to include high noise
pixels in its circumference, as well as the two point sources.

A circle can also be fitted though any set of three points (with the exception
of three collinear points), so we would also expect that the Hough transform of
an arbitrary field would contain a peak corresponding to each combinations of
three point sources. However, in practice many  such peaks would correspond to
very large circles, or circles with centres lying well outside the image
boundaries. That is, if the image contains $N$ point sources then the Hough
space will include $N \choose 3$ peaks. In the case of perfect point sources,
the magnitude of these peaks would be the sum of the amplitudes of the three
selected sources. In practice they will deviate somewhat from this due to the
non-ideal beam pattern.

As the magnitudes of the Hough peaks arising from point sources are dependent
on the source intensities, it is impossible to remove their effects with a
simple threshold. Instead it will be necessary to examine circles detected with
the Hough transform to determine whether they have arisen via this mechanism.
We have not yet examined this problem in detail, though it will be an important
component of any system intended to automatically process survey data using the
Hough transform.

Combinations of more point sources are of course also possible if they
happen to lie on a common circle, though this becomes increasingly unlikely as
the number of points increases.

The overall procedure proposed to find peaks in Hough space is as follows:
\begin{enumerate}
  \item Find the peak Hough transform amplitude for each value of $r$.
  \item Calculate how many standard deviations above the expected noise-only
        level $2 \pi r \hat{a}$ the maximum value of the Hough transform lies
        for each $r$.
  \item Search for local maxima in the resulting amplitude as $r$ varies.
        Such local maxima correspond to circles in the input image.
  \item Check for circles that obtain most of their votes from two (or
        more) point sources and remove them from the
        list of possible diffuse emission candidates.
\end{enumerate}
An example of this procedure is provided in Section \ref{sec:Testing_known}

\section{Results} 
\label{sec:Results}
We applied the CHT to images containing a variety of arc-like radio sources in
order to test features of the algorithm including filtering and robustness in
the presence of noise. The variant of the CHT used is described in
\citet{Hollitt_2009}, though the results described below are not dependent on
the particular implementation of the algorithm. Each of the three species of
source tested --- SNRs, tailed radio galaxies and giant radio relics ---
represent populations which will be present in next generation radio surveys.
It is expected that important science will be determined from the statistical
characterisation of such populations, making reliable automatic classification
an important goal.

\subsection{Response to non-diffuse structures}
\label{sec:Noise_Response}
We expect that the distribution of the Hough transform of a Gaussian noise
field $\sim\mathcal{N}\left( \hat{a}, s^2 \right)$ will be centered on $2 \pi r
\hat{a}$ with standard deviation $s \sqrt{2 \pi r}$. This allows us to
determine an appropriate threshold to assess the reliability of any proposed
circle detections.

Gaussian noise fields of $243 \times 243$ pixels were prepared with the pixels
values distributed with known mean and variance. The Hough transform was then
calculated to find the distribution of the magnitudes in Hough space.
For each value of $r$ the largest value in the Hough transform was found and
compared with the theoretical distribution.

\begin{figure}
  \centering
  \includegraphics[width=\linewidth]{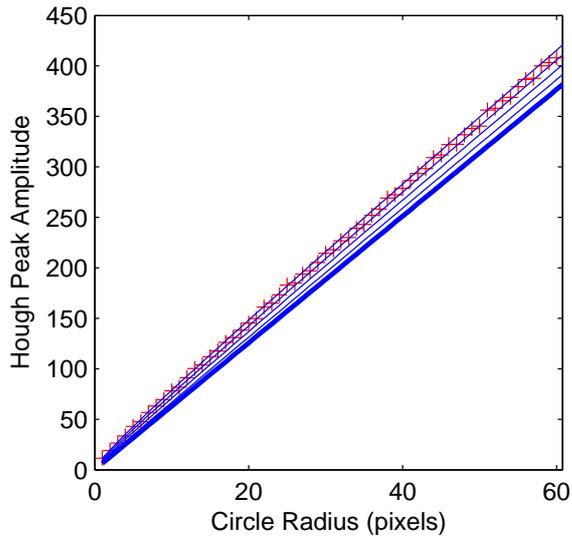}
  \caption{The red crosses indicate the largest value of the Hough transform,
  taken across the $243^2$ values for each target circle radius.
  The bold blue line indicates the expected value for the Hough
  transform of the random input field, with the thinner lines indicating
  progressive steps of one standard deviation in the expected value of
  the Hough transform.}
  \label{fig:Hough_noise}
\end{figure}

Fig \ref{fig:Hough_noise} shows the results of one such simulation, having the
input pixels distributed as $\sim\mathcal{N}\left( 1, 0.25 \right)$.
As can be seen the largest values for the Hough transforms typically lie
between three and four standard deviations from the expected value of the
transform. This is consistent with the expected extreme value of $243^2$
samples drawn from a normal distribution.

We therefore conclude that for the image size of $243^2$ pixels used throughout
this paper, a threshold of five times the input image standard deviation is
sufficient to exclude features in Hough space arising from image noise.

\begin{figure}
  \centering
  \includegraphics[width=\linewidth]{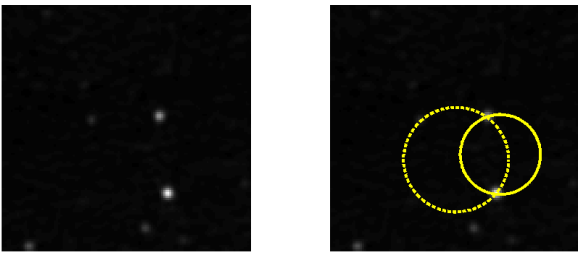}
  \caption{Response of the Circle Hough Transform to a collection of point
  sources. The centre of the solid circle lies between the two bright point
  sources. The dashed circle also passes through a third point source to the
  left of the image. }
  \label{fig:Hough_points}
\end{figure}

Figure \ref{fig:Hough_points} shows the result of applying the Circle Hough Transform to an
image from the Molonglo Galactic Plane Survey (MGPS) \citep{Green99}
containing no known diffuse structures, but a number of isolated point
sources. As discussed in \ref{sec:Detection}, the Hough transform results in a
peak corresponding to a circle centered roughly between the two brightest point
sources. It also finds a second circle that additionally passes through a third
point source.

\subsection{Supernova Remnants}
\label{sec:Testing_known}

When a massive star at the end of its life becomes a supernova, an energetic
shock wave is released from its surface. As the shock travels outward it
accelerates electrons in the interstellar medium to produce emission at radio
frequencies. In the simplest case the resulting emission would be spherically
symmetric as the spherical shock passes through a homogeneous interstellar
medium. However, where the shock wave is asymmetric, or when there is exiting
structure in the gas cloud surrounding the supernova, then more complicated
supernova remnants are produced \citep{Gaensler06, Orlando07}.

Our galaxy contains a large number of known supernova remnants (SNRs); the
result of supernov{\ae} occurring in roughly the last million years. The
Molonglo Observatory Synthesis Telescope Supernova Remnant Catalogue (MSC)
\citep{Whiteoak96} contains a list of 75 SNRs from the 843~MHz survey of the
Southern Galactic plane within the area 245$^{\circ} \leq l \leq 355^{\circ}$
in Galactic longitude and b~=~$\pm 1.5^{\circ}$ from the Galactic plane.
Statistical studies of the SNR population allows much more information to be
inferred than would be possible by examination of one SNR alone. In addition to
the valuable information gleaned on the average nature of the ISM surrounding
supernov{\ae}, statistical analysis shows that there is presently a significant
discrepancy between the total number of known SNRs and that expected from
stellar lifecycles \citep{Brogan06}. Comparison with the expected rate of
supernov{\ae} production from OB star counts, pulsar creation rates, iron
abundance and observation of supernova in other galaxies suggests that there
should be over 1000 SNRs detectable in the Galactic plane
\citep{Li91,Tammann94}. However, at present only $\sim$250 have been found
despite an increasing number of surveys of the region \citep{Green04,Brogan06}.
The detection of the population of SNRs is therefore an important aspect of
radio surveys of the Galactic disk.

The missing SNRs are believed to be those that are young and distant and those
which are old and nearby \citep{Green04,Brogan06}. In the former case the issue
is simply a lack of resolution in current surveys of the Galactic plane leading
to such small angular sized, bright sources being misclassified as point
sources. This issue will be readily addressed and resolved by the next
generation of radio telescopes such as the SKA. In the latter case however, the
SNRs will be both faint and large and if the present manual ``eyeballing''
techniques continue to be used  \citep{Whiteoak96, Filipovic02, Brogan06} they
are likely to remain elusive. Additionally, manual inspection is error prone
and often results in objects being missed in the initial data release
\citep{Whiteoak96,Green04}. Finally, while this approach has been adequate
(although not ideal) in the past, future generations of radio telescopes will
generate such a vast stream of data that manual inspection will not be
practical.

\begin{figure*}
  \centering\hbox{
   \includegraphics{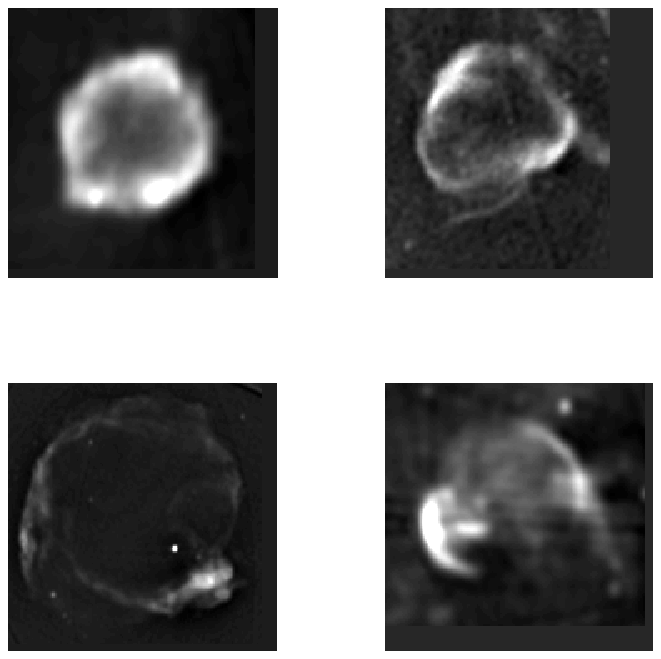}\hspace{2.2cm}
  \includegraphics{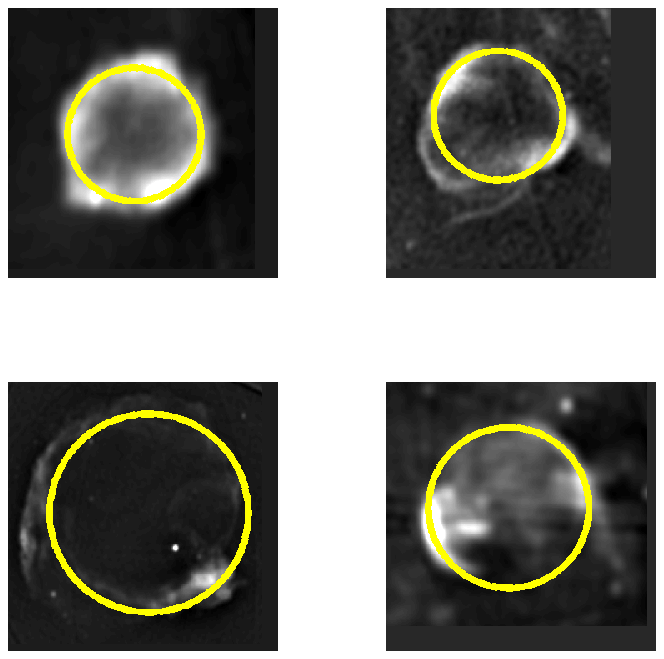}}
    \caption{Four supernova remnants taken from the MCS catalogue
\citep{Whiteoak96}.
           On the left panel subfigure a. is \mbox{G337.3+1.0},
           b. is \mbox{G302.3+0.7},
           c. is \mbox{G315.4-2.3} and
           d. is \mbox{G317.3-0.2}.
           On the right panel are the same four objects overlaid with circles
           that indicate the features detected via the Hough Transform
           of each object.}
  \label{fig:HT_results} 
\end{figure*}

We have selected a subset of four representative supernova remnants from the
MCS catalogue, as shown in Fig \ref{fig:HT_results}. This subset was selected
to range from a near complete ring (\ref{fig:HT_results}a), through two partial
rings (\ref{fig:HT_results}b and \ref{fig:HT_results}c) to a distorted example
(\ref{fig:HT_results}d). These four SNRs were intended to provide an increasing
challenge for the algorithm.

A CHT algorithm was performed on each of the four supernova remnants using the
procedure outlined in \citep{Hollitt_2009}. The four input data sets were each
padded to 243 pixels square. A Hough transform calculation is typically
preceded by an edge detection step to highlight the boundaries of objects.
However, supernova remnants are inherently shell-like, so we found the
edge-detection step to be unnecessary. Comparison of the transformation of the
raw and edge-detected data resulted in differences in only a few pixels in the
estimated SNR position and radius.

\begin{figure}
  \centering
  \includegraphics[width=\linewidth]{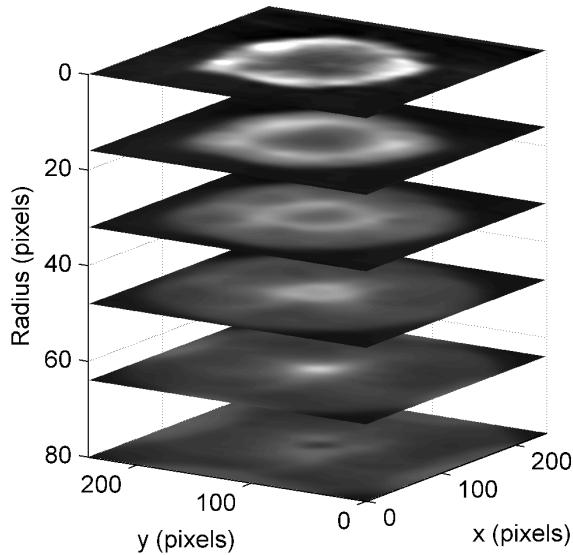}
  \caption{Slices through the Hough transform of \mbox{G337.3+1.0}. Slices are
  taken through several planes corresponding to a set of different possible SNR
  radii, $r=(16,32,48,64,80)$ pixels.}
  \label{fig:Hough_space}
\end{figure}

Figure \ref{fig:Hough_space} shows the Hough transform of one of the SNRs. For
clarity we selected the almost complete ring of G337.3+1.0. While the Hough
transform results in a full three-dimensional array this is difficult to present
graphically. The figure therefore shows the magnitude of the Hough transform
obtained for a small subset of the possible SNR radii. As can be seen for small
radii the intensity pattern is well distributed, indicating that there is no
strong peak for these radii and hence no small radius circles in the image.
However, in the $r=64$ plane we can see a strong peak toward the centre of the
field. This indicates that the input image contains a circle with radius
approximately $64$, which is located toward the centre of the image. For larger
radii we again see the intensity pattern is spread, so there are no
larger circles present.

\begin{figure}
  \centering
  \includegraphics[width=\linewidth]{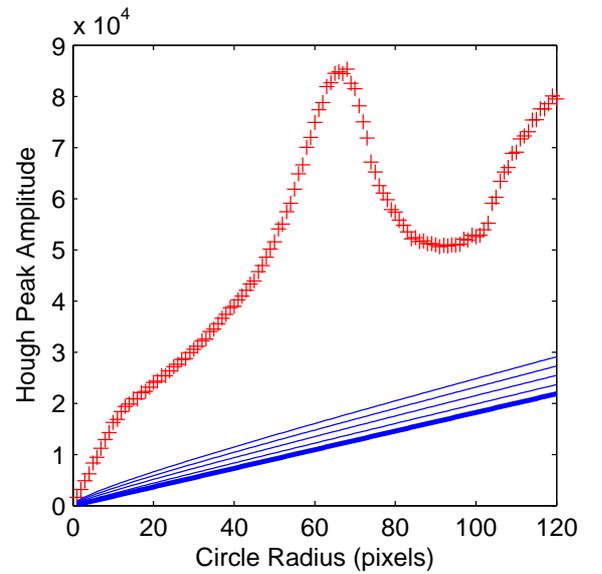}
  \caption{Maximum values of the Hough transform of \mbox{G337.3+1.0} as a
  function of radius are indicated by the red crosses. The thick lower line
  indicates the Hough transform expected from the image noise alone, with the
  increasingly higher lines indicating one standard deviation in the expected
  noise-only values.}
  \label{fig:G337_amplitudes} 
\end{figure}

The Hough transforms of the four SNRs were searched to find their maxima. As an
example of the procedure used, Fig \ref{fig:G337_amplitudes} shows the peak
value of the Hough transform of \mbox{G337.3+1.0} as a function of the target
circle radius. Also shown in the figure are the peak values expected if the
image contained noise alone. The figure clearly shows that the input image
contains features other than noise. 

\begin{figure}
  \centering
  \includegraphics[width=\linewidth]{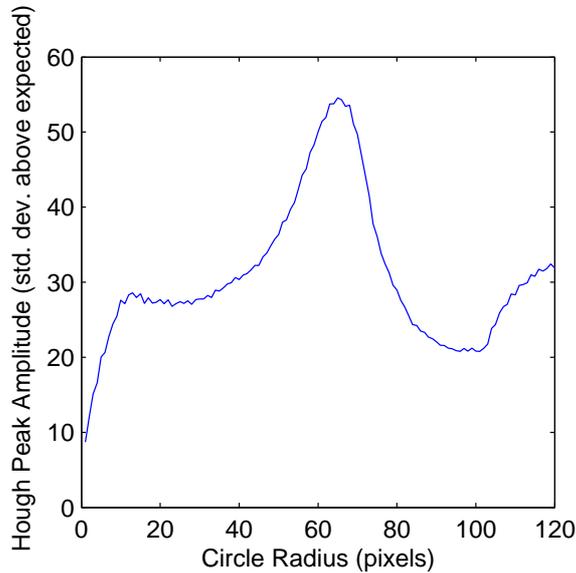}
  \caption{Maximum value of the Hough transform of \mbox{G337.3+1.0} as a
  function of radius. The amplitude is expressed as the number of standard
  deviations between the peak value and the value expected from a
  field consisting solely of Gaussian noise.} 
  \label{fig:G337_stddevs}
\end{figure}

The height of the peak above the noise-only level is then calculated in units
of standard deviations. that is, we calculate the statistical significance of
the Hough transform. The result of this calculation is shown in Figure
\ref{fig:G337_stddevs}. Local maxima in this graph indicate peaks in Hough
space, which can then be used to extract the location of circles in the
original image. In the case of Figure \ref{fig:G337_stddevs}, the prominent
peak at a radius of 65 pixels corresponds to the shell of the supernova
remnant. The local maxima around 10 pixels is associated with the circular
feature in the lower right portion of the SNR, as seen in Figure
\ref{fig:HT_results}a.

These peaks in the transforms correspond to the circle that is most consistent
with the input data in each case. Extraction of the coordinates of the peaks
therefore results in an estimate of the location and radius of the supernova
remnants. The effectiveness of the CHT in identifying the circular structures in the four
supernova remnants can be seen in Figure \ref{fig:HT_results}. In all four cases
the best fitting circles returned by the algorithm are in good agreement with
the radio data. 

\subsection{Bent-tail Radio Galaxies }
\label{BT}
\begin{figure}
  \centering
  \includegraphics[width=\linewidth]{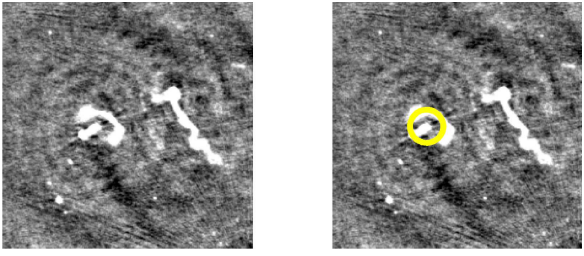}
    \caption{1.4 GHz image of a pair of head-tailed radio galaxies located
    in the centre of Abell 3125 \citep{mjh04}. The classic `C' shaped HT (SUMSS
    J032752-532613) has been correctly identified by the CHT over all of the
    other radio sources and artifacts in the image. }
  \label{fig:Head_Tail}
\end{figure}

Bent-Tail (BT) galaxies are Fanaroff-Riley class I/II \citep{fanaroffriley}
radio sources that consist of a core, coincident with the optical host galaxy,
and two tails of bent radio emission. The bending of the radio jets is believed
to result from the actions of ram pressure \citep{GunnGott} due to the relative
motion of the galaxy and its surrounding. In the classic case this results in a
`C' shaped radio object such as SUMSS J032752$-$532613 \citep{mjh04} seen on
the left of Figure \ref{fig:Head_Tail}, though more complicated structures such
as PKS J0327$-$532 seen on the right of the same figure do occur.

BTs have been found in the densest and most turbulent regions of galaxy clusters
in the local ($z<0.07$) Universe \citep{mao2009} at a rate of about 1--2 per
cluster. BTs are thought to act as signposts for over-densities in large-scale
structure, with associations between tailed-galaxies and clusters persisting out
to at least z $\sim$ 2  \citep{Dehgahn12a,Dehghan12b} making them an important
class of objects for science with future all-sky radio survey. 

One of the most important parameters to extract when using BT galaxies to obtain
physical information on the environment in which they are embedded is the radius
of curvature of the source \citep{freeland08,freeland10}. This technique is only
applied to wide-angle tailed sources (WAT) such as SUMSS J032752$-$532613, so it
is typically performed subsequently to the initial source detection phase. For
the test described in this section we wished only to find the radius of the
WAT, rather than extract information about the many wiggles in the tails of the
other BT. Consequently we imposed a constraint on the allowable circle radii
before running the transform. This demonstrates the ability of the CHT to search
for sources of only a certain angular size and to perform the simultaneous
science-driven detection and charactisation of the WATs. The
successful results are seen in the left of Figure \ref{fig:Head_Tail}.

As discussed in more detail in Section \ref{artifacts}, radio telescopes with
only a few elements suffer from gaps in the uv-plane which result in circular
artifacts in the image-plane. In the case of the image used here, the image,
from \citet{mjh04}, comprises of a mosaic of 4 pointings of 1.4 GHz data from
1.5 and 6 km configurations of the pre-CABB Australia Telescope Compact Array
(ATCA) which is known to have significant gaps in the distribution of baseline
lengths. The size of the resultant artifacts in the image plane is a function
of the antenna element spacing and is therefore known in advance allowing us to
employ filtering to prevent the CHT returning artifacts as sources. As next
generation radio telescopes will employ a larger number of elements, this is
likely not to be a major issue for future surveys.

\subsection{Radio Relics}

\begin{figure}
  \centering 
  \includegraphics[width=\linewidth]{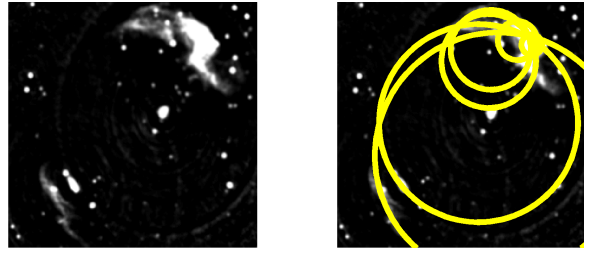}
    \caption{1.4 GHz Australia Telescope Compact Array (ATCA) mosaic image of
    the double relic cluster Abell 3667 \citep{mjh03}, this image covers
    approximately four square degrees on the sky and shows an example of one of
    the largest arc-like sources known. The CHT has identified the arc-like shape
    of the northern relic as well as bifurcations in the intensity structure
    of the source.}
  \label{fig:Relic}
\end{figure}

Radio relics are a rare class of large arc-like radio sources seen on the
periphery of galaxy clusters. They are believed to be the result of shockwaves
which occur when clusters of galaxies merge, the most powerful events since
the Big Bang. The physical size of these sources can be up to six million light
year across, such as in the northern relic in Abell 3667 seen in Figure
\ref{fig:Relic}. Because of their extremely low surface brightness they are
very difficult to detect with conventional pixel threshold based searches. As
in the case of faint SNRs, radio relics should be more easily detected with
algorithms utilising non-local information such as the CHT.

We tested the CHT on the field of Abell 3667 using 1.4~GHz ATCA data from
\citet{mjh03}. These data use five different configurations of the telescope
and so have good sampling of the uv-plane and hence a minimum of circular
artifacts. This particular image has had a taper applied in the uv-plane to
give a resolution of approximately one arcminute. The data contain a number
of extended sources including the double relics for A3667 that are located to
the top right and bottom left of the image, two BT galaxies in the centre of the
field and at the lower left above the relic and a number of unresolved point
sources. The CHT was employed without the filtering described in Section
\ref{BT} and the resultant detected features in Hough space are displayed on
the right side of Figure \ref{fig:Relic}.

Use of the CHT should isolate the location and scale of all circular features
in a set of radio data to provide a catalogue of possible sources that can be
assessed by an astronomer. In this case the algorithm correctly identifies the
curvature of the northern relic and gives the approximate cluster centre
(largest circle). It also correctly characterises an internal curved structure
in this relic (smallest circle). However,  the characterisation of the relics
is imperfect as Abell 3667 has an elliptical, rather than perfectly circular
structure. Also the middle circle is influenced by the bright BT galaxy just
north of the cluster centre which is unrelated to the relic. 

\subsection{Robustness to Noise}
\label{sec:Testing_noise}

To illustrate the robustness of the CHT to noise we contaminated the supernova
remnants in Figure \ref{fig:HT_results} with increasing amounts of Gaussian
noise and processed the resulting images until the SNRs could no longer
reliably be detected.

\begin{figure} 
  \centering
  \includegraphics[width=\linewidth]{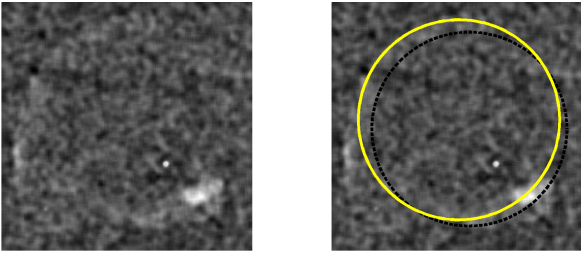}
  \caption{The input and results when the noise level in the supernova remnant
  from Figure \ref{fig:HT_results}c is increased until the Hough transform
  method is 50\% reliable. The solid line indicates the location of the SNR as
  estimated by the Hough transform. The dashed circle indicates the location of
  the SNR as predicted without additional noise added to the image before
  processing.}
  \label{fig:HT_noise}
\end{figure}

Figure \ref{fig:HT_noise} shows the supernova in Figure \ref{fig:HT_results}c
contaminated with additional noise until the Hough transform detector was 50\%
reliable. For this object the magnitude of the noise has been increased by a
factor of 10 from that in the original image. As shown in the figure, the
estimated location of the SNR in the contaminated image is slightly different
to that in the uncontaminated image. However, as we are primarily interested in
the detection of supernova remnants rather than their characterisation, we
do not regard this as a significant problem. As might be expected the error in 
the suggested location of the object typically decreases as the signal-to-noise 
ratio of the images improves.
 
\section{Discussion}
\label{sec:Discussion}
\subsection{Computational Issues}
Unfortunately the CHT is a computationally intensive algorithm, particularly
when used to characterise circles with unconstrained radii \citep{Yuen_1990}.
The CHT is therefore likely to prove too slow to directly process the vast
amounts of raw data that will be produced by the next generation of radio
telescopes such as the SKA.

Instead, we propose a two tiered approach to the detection of circular and
arc-like sources. The first stage will require the detection of regions of
extended radio emission (that is emission from any objects other than point
sources). This initial processing must be fast, but doesn't need to be very
accurate. Some variety of blob-detection algorithms would be suitable for this
purpose. A fair amount of work has been carried out by the astronomical
community on this problem and standard algorithms such as VSAD 
\citep{Condon98} and Duchamp \citep{Whiting11} exist. Ideally the blob
detection algorithm would additionally provide an estimate of the scale of a
possible source. However, existing astronomical application of blob detection
are poor at detecting broken large scale structures, so are likely to provide a
poor estimate of both the size and centre of most SNRs or double relics.

\subsection{Artifacts}
\label{artifacts}
Existing telescopes currently operating with $\nu \leq 3$~GHz necessarily use
a relatively sparse sampling of spatial frequencies, which can result in
circular calibration artifacts in their output data. Such artifacts would
confound our detection algorithm in its simplest incarnation. However, as
demonstrated in Section \ref{BT}, grating lobes of known size can be filtered
out before the final catalogue is returned. Additionally, many artifacts have
circular features combined with radial spikes, allowing a combined CHT and
linear Hough transform to be used to find and reject such sources. This would
have the additional benefit of being able to identify flawed regions in
an automatically generated image before a survey was
released.

\section{Conclusions}
\label{sec:Conclusion}
We have examined the Circle Hough Transform as a detection and characterization
method for finding circular or arc-like sources in astronomical data. We have
tested the algorithm on MGPS and ATCA images of a range of sources including
supernova remnants, tailed radio galaxies and radio relics which all represent
scientifically important classes of radio sources for the next generation
of radio telescopes. We have shown that the technique is successful at
recovering such sources and providing physical information for them in an
automated way.

At present thresholding algorithms such as Duchamp \citep{Whiting11} are able to
produce catalogues of radio sources based on the requirement of having a
certain number of contiguous pixels over a certain signal-to-noise level. These
algorithms will usually return either the position of the brightest pixel or
the geometrical centre of the source as the central coordinate. This approach
will work adequately for the majority of symmetrical, unbroken sources.
However, for arc-like sources such techniques will not produce the correct
coordinates making automatic cross-matching to other wavelengths difficult (as
is currently the case). Additionally, with broken sources such as SNRs and
double relics, such techniques will report multiple, unrelated diffuse sources
(all with the wrong central coordinates) rather than identify a single source.

Because it uses non-local information, the CHT is able to work with broken
sources and is capable of detecting sources that have much lower
signal-to-noise ratios than other algorithms. The CHT can therefore be
expected to do a superior job of both detecting and characterizing some
classes of sources, particularly those that are very diffuse.

However, as the CHT is computationally expensive, it is unlikely to be sensible
to run the algorithm over the entire sky for an all-sky survey such as EMU
\citep{Norris11}. We propose that it would be better to use a conventional
``blob detection algorithm" to preprocess the data before and then apply the
CHT in certain parts of the sky only. These regions could either be selected
based on criteria from the results of the conventional blob detection, or as
regions where such low-surface brightness arc-like sources are expected such as
in the Galactic plane or around galaxy clusters. This tiered approach is likely
to be the most efficient method for generating a robust catalogue of
diffuse radio sources and will be the subject of further experimentation.

There are several improvements that could be made to the current CHT. Firstly,
the robustness of the method can be improved by the development of a more
effective method for searching the Hough transform for peaks. Simply finding the
maximum value as we have done here is unlikely to be the optimal technique.
Secondly, as seen in some SNRs and double relics, the objects in question are
elliptical rather than circular. The Hough transform could be extended from the
current circular implementation to an elliptical version. However this increases
the number of parameters required and hence would result in a five dimensional
Hough space which would considerably increase the computation complexity.

As has been noted, the Circle Hough Transform produces apparent circles
as a result of point sources. If the Hough transform is to prove effective at
automatic diffuse source detection, it will need to be combined with
additional processing to remove the effects of point sources. Once this has
been done the effectiveness of the combines detection system can be assessed by
using it on a catalogue containing known diffuse structures.   

A catalogue of known supernova remnants has been manually compiled from the
second Molonglo Galactic Plane Survey (MGPS2) \cite{Murphy07}, allowing
ready comparison with the results of the automatic detection. Use of the CHT on
such a dataset will allow the determination of the false negative and
false-positive detection of the technique. This comparison also admits the
possibility of the detection of as-yet undiscovered supernova remnants and in
particular those in the class of being old and faint which have been predicted
via several independent indicators  but remain, as yet, undetected. Thus, in
addition to characterising the false-positive and false-negative error rates,
performing the CHT on the MGPS2 could address one of the current problems in
understanding the so-called ``Galactic Ecology''.

\section*{Acknowledgement}
The MOST is operated by The University of 
Sydney with support from the Australian Research Council and the Science 
Foundation for Physics within The University of Sydney. The Australia Telescope 
is funded by the Commonwealth of Australia for operation as a National Facility 
managed by CSIRO.

\bibliographystyle{plainnat}

\end{document}